\documentclass[aps,onecolumn,notitlepage,10pt]{revtex4-1}

\usepackage{graphicx}  
\usepackage{bm}        
\usepackage{amssymb,amsfonts,amsmath}   
\usepackage{color}  
\usepackage{lipsum}
\usepackage{upgreek}
\usepackage{mathtools}
\usepackage{comment}

\renewcommand{\d}{\mathrm{d}}

\renewcommand{\d}{\mathrm{d}}
 
\newcommand{\x}{{\boldsymbol x}}

\definecolor{light}{rgb}{0.81,0.81,0.81}
\newcommand{\0}{\textcolor{light}{0}}
\DeclarePairedDelimiter\floor{\lfloor}{\rfloor}
\allowdisplaybreaks

\begin{document}

\title{The role of tumbling frequency and persistence\\in optimal run-and-tumble chemotaxis}

\author{Julius B. Kirkegaard and Raymond E. Goldstein}
\affiliation{Department of Applied Mathematics and Theoretical Physics, Centre for Mathematical Sciences, \\ University of Cambridge, Wilberforce Road, Cambridge CB3 0WA, United Kingdom}

\date{\today}

\begin{abstract}
{One of simplest examples of navigation found in nature is run-and-tumble chemotaxis.
Tumbles reorient cells randomly, and cells can drift toward attractants or away from repellents 
by biasing the frequency of these events.
The post-tumble swimming directions are typically correlated with those prior, as measured by the variance
of the reorientation angle distribution.
This variance can range from large, in the case of bacteria, to
so small that tumble events are imperceptible, as observed in choanoflagellates.
This raises the question of optimality: why is such a range of persistence observed in nature?
Here we study persistent run-and-tumble dynamics, focusing first on the optimisation of the linearised chemotactic response within the
two-dimensional parameter space of tumble frequency and angular persistence.  Although an optimal persistence does 
exist for a given tumble frequency, in the full parameter space there is a continuum of optimal solutions.
Introducing finite tumble times that depend on the persistence can change this picture,
illuminating one possible method for selecting tumble persistence based on 
species-specific reorientation dynamics.
Moving beyond linear theory we find that optimal chemotactic strengths exist,
and that these maximise reaction when swimming in a wrong direction,
but have little or no reaction when swimming with even the slightest
projection along the chemoattractant gradient.}
\end{abstract}

\maketitle

\section{Introduction}

Chemotaxis, the ability to navigate concentration fields of chemicals,
is a ubiquitous feature of the microscopic world.
Performed by uni- and multicellular organisms alike, 
it represents one of the simplest forms of navigation.
Within this simplicity, various strategies exist.
For instance, certain spermatozoa measure chemoattractant
gradients by swimming in helical trajectories and bias the helical axis to move directly 
towards the chemoattractant \citep{Friedrich2007, Jikeli2015}.
Green algae can swim towards light-intense regions by measuring light source directions
as they rotate around their own swimming axis \citep{Yoshimura2001, Drescher2010},
but also bias their navigation by switching between synchronous and anti-synchronous beating of their flagella \citep{Polin2009}.
The slime mould \textit{D. discoideum} is large enough to measure directly the spatial concentration gradients in cAMP, along which it navigates \citep{Bonner1947}.
The epitome of chemotaxis is perhaps the run-and-tumble of certain
peritrichously flagellated bacteria such as \textit{E. coli} \citep{Berg1972}.

\begin{figure}[t]
\centering
\includegraphics{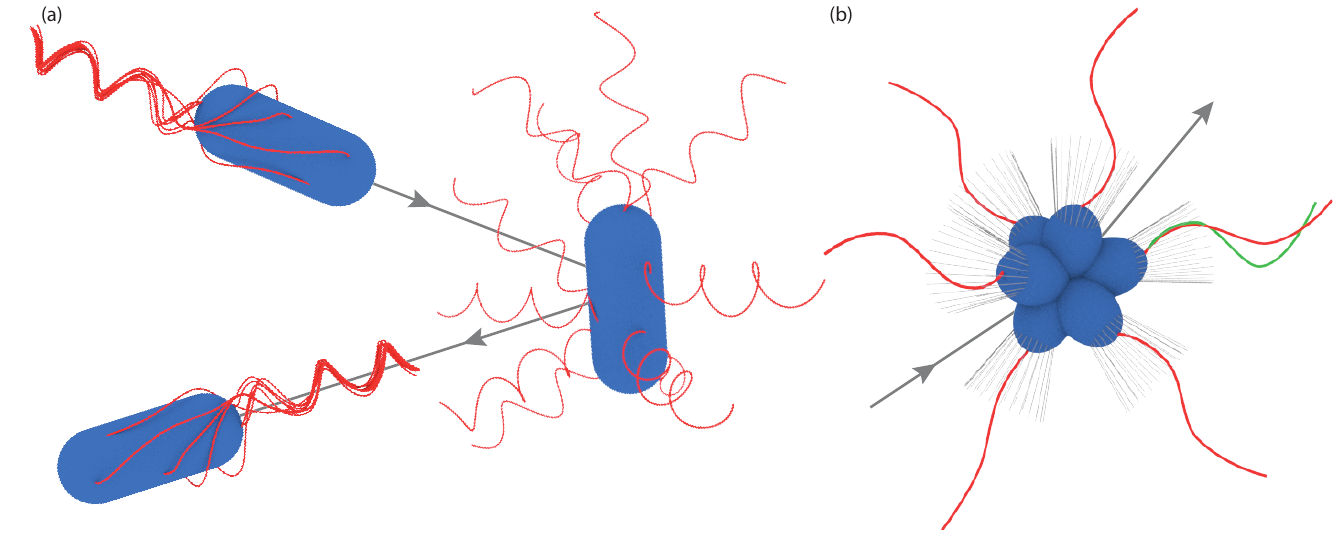}
\caption{Run-and-tumble. (a) Swimming and tumbling of a peritrichous bacterium. During 
swimming the flagella rotate counter-clockwise and form a bundle resulting in a run. Clockwise 
rotation one or more of the flagella breaks the bundle and results in a tumble that reorients the cell. (b) Choanoflagellate colony reorientation event. Each cell's flagellum beats independently of
the others. A change in the beating dynamics of one cell (green flagellum) can cause a small reorientation of the colony as a whole.}
\label{fig:runtumble}
\end{figure}

Run-and-tumble motion is comprised of approximately straight lines (runs) interrupted by
reorientation events (tumbles), as shown in Fig. \ref{fig:runtumble}.
For example, in peritrichous bacteria, the helical flagella rotate counter-clockwise and form a coherent bundle during swimming.
A tumble is induced when (some of) the flagella reverse their rotational direction and the bundle is 
disrupted (Fig. \ref{fig:runtumble}a).
This creates a large, transient, reorientation.
Navigation along a gradient of chemoattractant becomes possible if the frequency of
tumbling is biased in response to the chemoattractant distribution. 
This is a type of \textit{stochastic} navigation in the sense that the organisms that perform it
do not swim directly in the desired direction, but rather in a random direction and later decide
whether such a turn was correct.

\begin{figure}[t]
\centering
\includegraphics{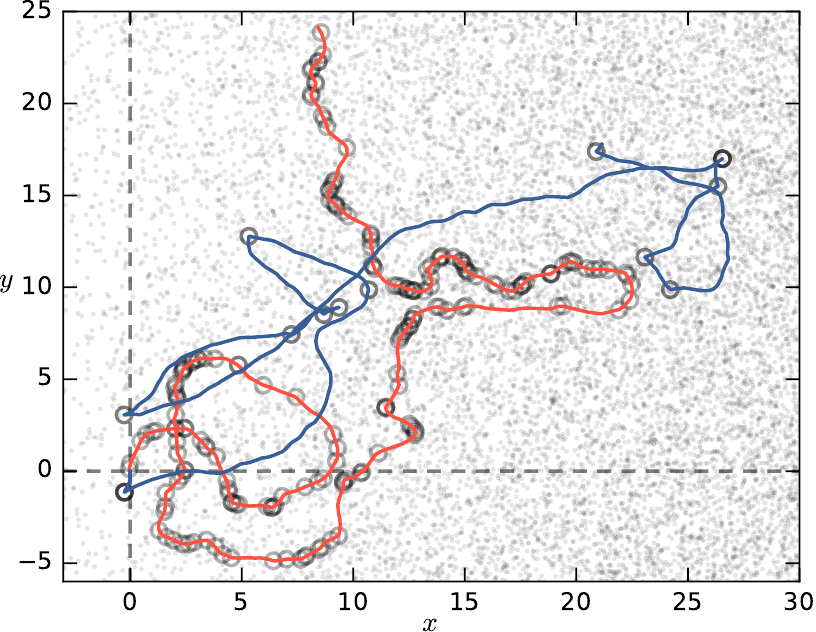}
\caption{Run-and-tumble trajectories. Both simulated trajectories drift to the right, starting from the origin. Full tumbles are shown in blue ($k=0, \, \lambda_0 = 0.1$) and persistent but frequent tumbles in red ($k=10, \, \lambda_0 = 2.0$). Circles indicate tumbles. Shared parameters: $D_r = 0.1, \, \beta = 1/2, \, \gamma=1.$}
\label{fig:chemotaxis_strategies}
\end{figure}

The tumbling frequency is modulated through measurements of the variation in concentration of chemoattractants, illustrated by the background of Fig. \ref{fig:chemotaxis_strategies}.
In an idealised scenario, the reorienting tumbles result in unbiased new directions, 
uniformly chosen from the unit sphere, 
but this is not typically the case.
Instead, a persistence with the previous direction is present \citep{Berg1972}.
In fact, for some species, the individual reorientations are so small that they are hardly observable.
This is the case in colonies of choanoflagellates \citep{Kirkegaard2016a},
within which the flagella beat independently \citep{Kirkegaard2016b}
and a reorientation event may simply arise from slight modulation of the
beating of a single flagellum (Fig. \ref{fig:runtumble}b).
These smaller tumbles, or directionally persistent tumbles,
add up to a smoother swimming while still allowing navigation.
Fig. \ref{fig:chemotaxis_strategies} shows two realisations of run-and-tumble swimming.
In blue is the case of full-reorientation tumbles and in red is very persistent tumbles occurring with higher frequency.
Over long time-scales both of these swimmers perform random walks biased in the direction of the chemoattractant signal.

A strong theoretical understanding of chemotaxis exists \citep{Tindall2008a, Tindall2008b}, 
including the filtering of chemoattractant signals
to which the cells react \citep{Segall1986, Celani2010},
the fundamental limits of measurement accuracy of such signals \citep{Mora2010}
and the limits they impose on navigation \citep{Hein2016}.
Theories of chemotaxis are typically developed in
the weak-chemotaxis limit \citep{Celani2010, Locsei2007, Locsei2009, Reneaux2010, Mortimer2011},
the linear theory of which provides accurate explanations of many experimental observations.
Theory \citep{Locsei2007} and simulation \citep{Nicolau2009} of chemotactic bacteria have also showed
that for otherwise equal chemotactic parameters, directional persistence of tumbles,
as observed in experiments, can lead to enhanced chemotaxis.

This raises a more general question: could the effect of changing one parameter, such as 
directional persistence, be compensated by
simultaneously changing another?
Here, we address this question of global optimality, and examine effects that lead to
the existence of optima.
For example, in linearised theories, the drift velocities for large chemotactic strength
and for steep gradients can become unbounded, and thus the evaluation of one effect is
done at fixed chemotactic response.
But microorganisms do not have the restrictions that come with choosing theories that are analytically tractable.
In real systems, the drift velocity will be limited (trivially) due to the finite swimming speed
of the organisms and (more importantly) by uncertainties of measurements in noisy environments
combined with diffusion.
Throughout this study we optimise for the performance of a single organism, neglecting
population effects \citep{Peaudecerf2015}.

\section{Model}
The model of chemotaxis used here assumes that organisms determine concentration gradients by
comparing their concentration measurements at different times as they move through the
medium, rather than detecting gradients over
their own body, as is possible for organisms considerably larger than bacteria \citep{BergPurcell}. 
To be precise, we assume that as a cell swims it measures only the local chemoattractant 
concentration $c(\x, t)$ at its present position $\x$.
Moreover, in this section the concentration is taken to be linear in position, 
$c(\x) = c_0 + \alpha \, x$,
allowing the notation
$c(t) = c(\x(t))$ for a given trajectory $\x(t)$.
Cells are thought to store the history of these measurements,
and use this to bias their tumbling frequency $\lambda$.
In the present model, this is embodied by the relationship $\lambda = \lambda_0 \, \floor{1 + q}$,
where $\floor{\cdot} = \max(0, \, \cdot)$.
Here, $q$ is the biaser, determined by a linear convolution of $c$,
\begin{equation}
q(t) = \int_{0}^\infty c(t-t') \, \kappa(t') \, \d t'.
\label{q_eqn}
\end{equation}
We take the kernel to be one studied previously and which 
corresponds well to experimental measurements \citep{Segall1986, Celani2010},
\begin{equation}
\kappa(t) = \frac{\beta \gamma^2}{\alpha \, v} 
e^{-\gamma t} \left[\frac{\left(\gamma \, t\right)^2}{2} - 
\gamma \, t\right],
\end{equation}
where $v$ is the swimming speed and $\gamma$ is the memory time scale of past measurements.
The normalisation is chosen such that $\max |q| = \beta$ in the absence of noise,
and hence $\beta$ solely specifies the chemotactic strength.
The kernel satisfies $\int k(t) \, \d t = 0$
which gives perfect adaptation to any background chemoattractant concentration.
This criteria arises naturally from maximising the minimum chemotactic efficiency over 
all chemoattractant profiles \citep{Celani2010}.
In particular, this is an important feature that will not arise from
maximising drift velocity alone and which we thus impose \textit{a~priori} here.

We consider cells swimming in two dimensions in an instantaneous direction $\theta(t)$ with velocity $v$, and discuss the three-dimensional case in Appendix \ref{app:3d}.
This direction is modulated by both rotational diffusion as $\d \theta = \sqrt{2 D_r} \, \d W$, where 
$W$ is a standard Wiener process, and by tumbles, the size of which are chosen from a von-Mises 
distribution with parameter $k$, $p(\Delta \theta) = \exp(k \cos(\Delta \theta) )/2 \pi I_0(k)$, where 
$I_n$ are the modified Bessel functions of the first kind. Thus $k$ specifies the persistence of the 
tumbles, $k=0$ corresponding to full tumbles.

\section{Measurement Time-scale}
If the time $1/\lambda_0$ between tumbles is too long compared to the rotational diffusion time
$1/D_r$, the trajectories will be reoriented by rotational
diffusion and the organism will have lost the ability to bias its motion in any useful way.
Thus, the biasing of tumbles must outcompete rotational diffusion and we expect
$\lambda_0 \gtrsim D_r$.
In the absence of measurement noise, and if the organism can make instantaneous measurements 
($\gamma \rightarrow \infty$),
increasing the chemotactic strength will monotonically increase the chemotactic drift,
and in the limit $\beta \rightarrow \infty$ chemotaxis becomes perfect, despite the hindering effects of rotational diffusion.
But, we emphasise that this is only possible in the absence of measurement noise.
Here, in contrast, we are interested in the noise-limited situation, and with noise comes another time scale, that over which accurate measurements can be made (see Appendix \ref{app:1d} for
a simple lattice calculation illustrating this point).

To illuminate this situation we perform simulations in which cells are placed in a constant gradient (linear increase) of discrete chemoattractants.
In a periodic $2L \times 2L$ box, $N$ molecules are placed, decreasing linearly in concentration from $x=0$.
This is achieved by choosing each molecule's position as $x = L \sqrt{|U_1|} \, \text{sign}(U_1), \, y = L U_2$, where $U_i$ is uniformly distributed on $[-1,1]$.
$c(t)$ is then defined to be the number of molecules within a cell's area.
This can be evaluated efficiently by storing the molecules 
in a $k$-D tree, allowing for fast simulations with billions of molecules.

\begin{figure}[!t] 
\centering
\includegraphics{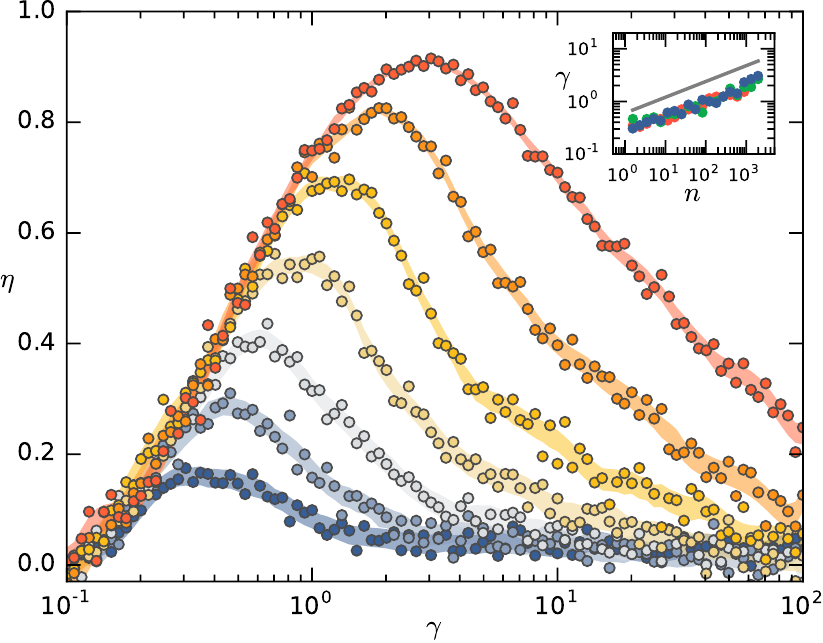}
\caption{Chemotactic efficiency as a function of $\gamma$. Each curve corresponds to different concentration levels; at the lowest concentration (blue data) each cell senses 
on average $n \sim 1.5$ molecules, while at the highest (red), $n \sim 2000$ are sensed.  Shaded background indicates standard error of the simulations. Inset shows optimal $\gamma$ as a function of $n$ for $\lambda_0 \in \{0.1, 1.0, 5.0 \}$. $D_r = 0.1, \, \gamma=1$.}
\label{fig:varygamma}
\end{figure}

For the purposes of the present discussion, we define the 
chemotactic efficiency proportional to the average value of
the concentration experienced by the organism in steady state
\begin{equation}
\eta \propto \langle c \rangle = \int \! c(\x) P(\x) \, \d \x ,
\end{equation}
where $P(\x)$ is the steady-state probability distribution.
The normalisation is chosen such that $\eta = 1$ corresponds to perfect chemotaxis.

Fig. \ref{fig:varygamma} shows $\eta$ as function of the measurement rate $\gamma$
for various molecular concentrations.
The curves clearly reveal the existence of an optimal $\gamma$ for each choice of
the (average) number of molecules sensed.
Choosing $\gamma$ too low means slow reaction, 
but with $\gamma$ too high the organism does not have time to make an accurate measurement
before previous information is forgotten.
Varying the chemoattractant concentration (but not the gradient) shifts the optimal $\gamma$.
At higher concentrations, the measurement noise is lower \citep{Mora2010},
and thus less time is needed to make an accurate measurement.
The inset of Fig. \ref{fig:varygamma} shows how the optimal $\gamma$ varies with the concentration,
and further shows that, within the resolution of our simulations, this optimum is independent of the base tumbling frequency $\lambda_0$.
This independence means that we can fix $\gamma$ to its optimal value without specifying the value of $\lambda_0$.

In this noise-dominated regime, $\gamma$ is thus set by the chemoattractant concentration.
If cells are kept in a chemostat with fixed concentration and gradient,
as is the case considered here, 
the optimal $\gamma$ is thus indicative of the underlying noise levels.
In the following sections we fix $\gamma$, thus implicitly defining the noise levels.
The goal then becomes to find the optimal choice of the remaining parameters for a given $\gamma$.
Our approach ignores spatial variations in noise,
but conclusions made are confirmed by checking them against
the full simulation setup used in this section.

\section{Tumbling Frequency \& Persistence}
In earlier theoretical work, persistence of tumbles has been shown to enhance the chemotactic drift 
velocity \citep{Locsei2007,Nicolau2009}.
Possible rationalizations for this effect include the idea of information relevance;
for persistent tumbles, the gradient information (stored in $q$, Eq. \ref{q_eqn}) remains more relevant
than for full tumbles, where a completely random direction is chosen.
It has also been shown that an optimum base tumbling frequency $\lambda_0$ exists \citep{Celani2010}.
Intuitively, in the low-noise limit, this optimum should be set by the rotational diffusion constant $D_r$,
in order to dominate rotational diffusion but not hinder drift.
Intuitively, one expects that introducing persistence, which results in smaller angular deflections 
from tumbles, would shift the optimal tumble frequency to higher values.
So while it is clear that persistence can increase the chemotactic drift for a given base tumble frequency,
it is not clear what the effect is if variations in $\lambda_0$ are also allowed.

\begin{figure}[!t]
\centering
\includegraphics{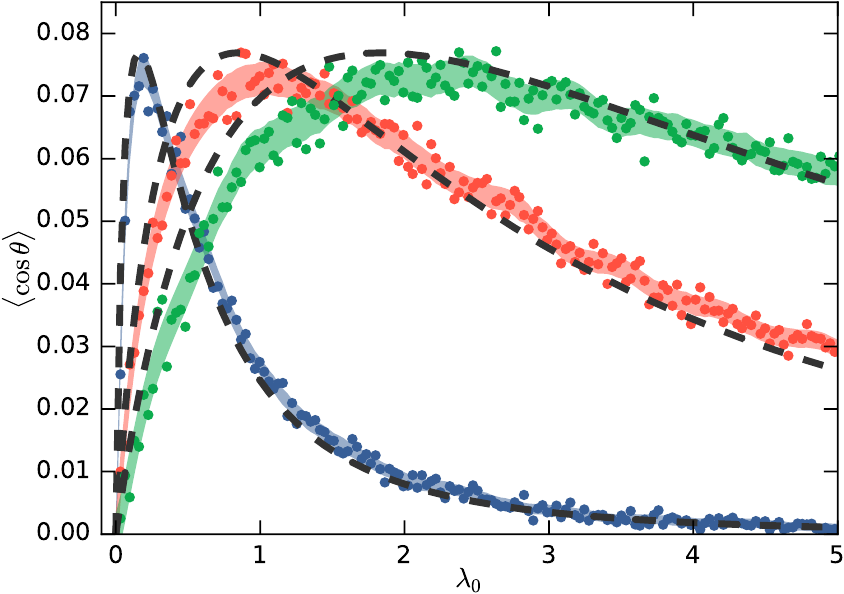}
\caption{Drift efficiency as a function of tumble frequency.   Data from direct simulation of
full model are shown for $k=0$ (blue), $k=3$ (red), and $k=6$ (green). Each data point is the result of 10,000 simulations and shaded background indicates standard error, with $D_r = 0.1, \, \beta = 1/2, \, \gamma=1$.  Linearised theory is indicated by dashed lines.}
\label{fig:vary_k}
\end{figure}

To study this, we simulated cells performing chemotaxis in a constant gradient for various persistence parameters $k$, while varying $\lambda_0$.
The results shown in Fig. \ref{fig:vary_k} confirm the intuition outlined above;
for large base tumbling frequency $\lambda_0$,
increasing the persistence $k$ leads to increased chemotactic drift, as previously found.
However, for low values of $\lambda_0$ the opposite effect is found.
There is thus a trade-off between frequency and persistence of tumbles.

To gain further insight we study the relevant Fokker-Planck equation.
As shown previously \citep{Celani2010}, the dynamics of the biaser $q$
can be made Markovian by introducing three internal variables (moments of $c$) 
\begin{equation}
m_j = \int_{-\infty}^t 
e^{-\gamma (t-t')} (t-t')^j c(t') \, \d t',
\end{equation}
which obey a coupled system of differential equations $\partial_t m_j = c(t) \delta_{j0} -\gamma m_j + j m_{j-1}$.
It follows that our system can be fully described by a Fokker-Planck equation for a distribution
function $P(x,\theta, \{ m_j \}, t)$
\begin{align}
\frac{\partial P}{\partial t} + v \cos \theta \, \frac{\partial P}{\partial x}  = & \, \, D_r \frac{\partial^2 P}{\partial \theta^2} + \lambda_0 \floor{1 + q(t)} \left[\int \, \frac{e^{k \cos(\theta - \theta')} }{2 \pi I_0(k)} P(\theta') \, \d \theta' - P \right] \\ \nonumber
&- \sum_{j} \partial_{m_j} [ \delta_{j,0} \, c(x) \, + j m_{j-1} - \gamma m_j ] \, P,
\end{align}
where 
\begin{equation}
q(t) = \frac{\beta \gamma^2}{\alpha \, v} \left(\frac{1}{2}\gamma^2 m_2 - \gamma m_1\right)~.
\end{equation}

We begin by solving this system for small $\beta$.
Later we will argue that our conclusions remain qualitatively correct also for large $\beta$.
In steady state, we find (Appendix \ref{app:lindrift})
\begin{equation}
\langle \cos \theta \rangle = \frac{\beta \gamma^3 \lambda_k}{2(D_r + \lambda_k)(D_r + \gamma + \lambda_k)^3} + \mathcal{O}(\beta^2),
\label{eq:lineardrift}
\end{equation}
where $\lambda_k = \lambda_0 \left[1 - I_1(k)/I_0(k) \right]$.
Note that the only place $\lambda_0$ enters is through the quantity $\lambda_k$, which
has the optimal value  
\begin{equation}
\lambda_k^*  = \frac{1}{3} \left(\sqrt{4 D_r^2 + 3\gamma D_r} - D_r\right)~.
\end{equation}
From this fact, we conclude that the trade-off between tumbling frequency and tumble persistence is perfectly
balanced; changes in $k$ can be precisely compensated by changes in $\lambda_0$.

Fig. \ref{fig:vary_k} shows how this small-$\beta$ result accurately matches the full numerical
results even for $\beta = 1/2$.
So while persistence can lead to enhanced chemotaxis, we find that this has nothing 
inherently to do with the persistence of the tumbles themselves, as the same increase can be achieved
simply by lowering the base tumbling frequency.

With constant $\lambda_0$, letting $k \rightarrow \infty$ results in negligible drift.
For large $k$, $\left[1 - I_1(k)/I_0(k) \right]^{-1} \sim 2k$.
Thus we see that a continuous version of run-and-tumble \citep{Kirkegaard2016a} emerges
in the limit $k \rightarrow \infty$ if $\lambda_0$ is scaled linearly with $k$,
and we conclude that such a strategy is equally optimal to any other persistence
of tumbles with the correct choice of tumble frequency.
These results arise because we allow $\lambda_0$ to be chosen independently of $\gamma$.
Without persistence, chemotaxis is optimised for $\lambda_0$ and $\gamma$ of similar order.
For cells with large persistence, however, optimisation leads to $\lambda_0$ much larger than $\gamma$.

The expansion $P(\theta, \cdots) = a_0(\cdots) + a_1(\cdots) \cos \theta$ captures the steady state distribution well,
and the form of $\lambda_k$ can only change if higher order Fourier modes become important.
This is not case even in the high $\beta$ regime,
and so these conclusions are also valid there.
Second order effects such as small dependencies of the optimal $\gamma$ on $\lambda_0$ and $k$ could
also perturb the result of perfect trade-off between tumble frequency and persistence.
Furthermore, although we only considered the steady state here,
the conclusions apply to the transient behaviour of the system.
Our conclusions also hold in three dimensions as demonstrated in Appendix \ref{app:3d}.

Real bacteria are observed to have an angular distribution of tumbles with a non-zero 
mode \citep{Berg1972}.
To model this, we consider the reorientation distribution $g(\theta, \theta') = \sum_{\pm}e^{k \cos(\pm \mu + \theta - \theta')} / \penalty 0 4 \pi I_0(k)$.  This results in the substitution $I_1(k)/I_0(k) \rightarrow I_1(k)/I_0(k) \cos \mu$ in Eq. \eqref{eq:lineardrift} (see Appendix \ref{app:meantumble}), leaving unchanged our conclusions.
If the turns are biased in one direction (\textit{e.g.} turning more clockwise than counter-clockwise),
such that $g(\theta, \theta') = e^{k \cos(\mu + \theta - \theta')} / 2 \pi I_0(k)$, the efficiency can surpass that of unbiased cells.
In this case the optimum strategy involves cells that continuously rotate,
modulating their rotation speed as they swim (Appendix \ref{app:biastumble}).
While this is interesting behaviour, such a bias is a 2D phenomenon,
although a related optimality may exist in 3D.

\begin{figure*}[!tb]
\centering
\includegraphics{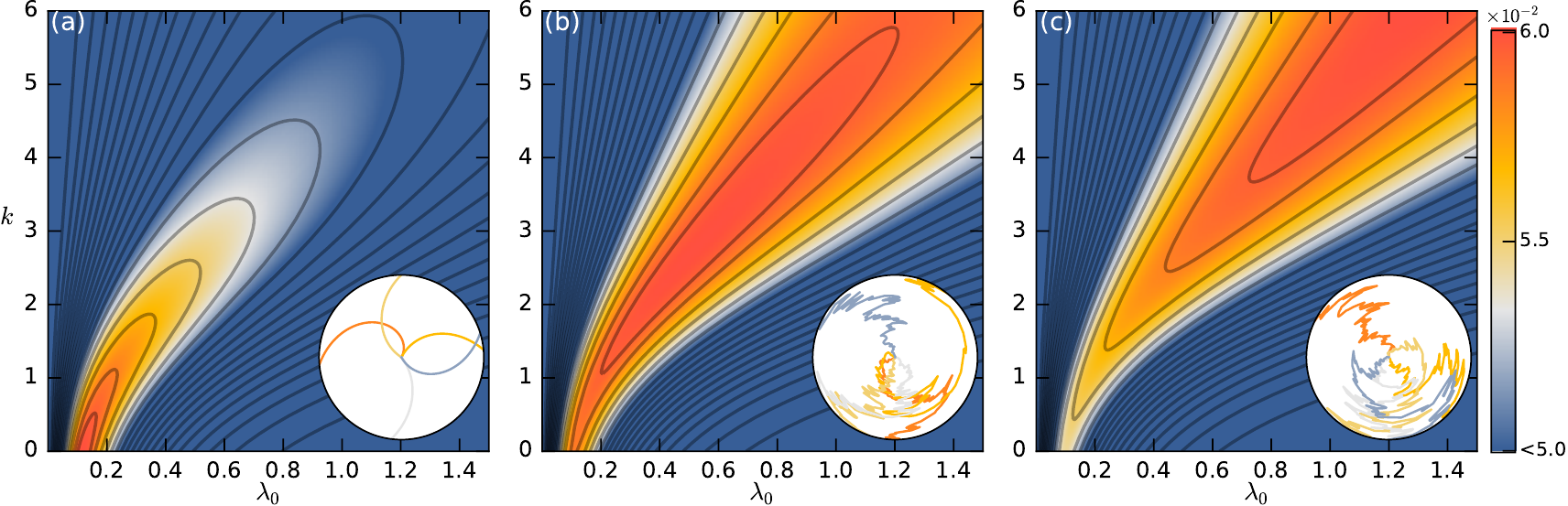}
\caption{Drift efficiency $\langle \cos \theta \rangle$ with finite tumbling time. Panels correspond to different tumble time exponents: (a) $\alpha = 1.0$, (b) $1.7$, and (c) $2.0$. Colour scale shows the top 20\% range of drift velocities. Insets show examples of reorientation trajectories for with exponent $\alpha$, the angle on the circle indicating the orientation and the radial distance indicating time. Common parameters: $D_r = 0.1, \, \beta = 1/2, \, \gamma=1,\, \tau_0 = 1$.}
\label{fig:finite_tumble_time}
\end{figure*}

The fact that no single persistence value is globally preferable
fits well with the experimental variations seen between biological species.
The question still remains, however, if there are other effects that could induce a preferred
tumble persistence.
So far we have assumed the tumbles to be instantaneous.
Including a finite tumble time can change the conclusions.
In particular, since the optimal tumbling frequency for persistent tumbles is large, 
adding a constant time for each tumble results in large amounts of time in which no 
chemotactic progress is made, hence disfavouring persistence,
On the other hand, one would expect a persistent tumble to take less time than a full tumble.
The average tumble time $\langle \tau \rangle$ should depend on the average angle turned.
The precise form of this dependence will change with reorientation method.
If the tumbling rotation is ballistic, the mean reorientation time should be proportional to the mean angle turned.
If, on the other hand, the cell relies on a diffusive method (which includes simply not swimming),
the reorientation time will be proportional to the mean of the squared angle.
We parametrise this with the exponent $\alpha$, with $\alpha  = 1$ for ballistic reorientations and
$\alpha =2$ for diffusive and a mixture for values in-between.
The mean tumbling time is thus
\begin{equation}
\langle \tau \rangle = \frac{\tau_0}{\pi I_0(k)} \int_{0}^\pi \! \delta^\alpha e^{k \cos \delta}\, \d \delta.
\label{eq:tumble_time_scaling}
\end{equation}
The insets of Fig. \ref{fig:finite_tumble_time} show trajectories for ballistic 
diffusive and intermediate exponents.
For small chemotactic strength it is easy to incorporate this effect.
The fraction of time spent swimming will be $1/(1+\langle \tau \rangle \lambda_0 )$, so we find
\begin{equation}
\langle \cos \theta \rangle \rightarrow \frac{1}{1 + \langle \tau \rangle \lambda_0 } \frac{\beta \gamma^3 \lambda_k}{2(D_r + \lambda_k)(D_r + \gamma + \lambda_k)^3}.
\end{equation}
Crucially, $\lambda_0$ now appears alone,
and we thus expect a global optimum to appear.
Fig. \ref{fig:finite_tumble_time} shows $\langle \cos \theta \rangle$ evaluated for various exponents.
For ballistic ($\alpha=1$) we find that full tumbles ($k=0$) are optimal.
For diffusive $\alpha=2$, the continuous dynamics ($k \rightarrow \infty$) become optimal.
In-between, as shown in Fig. \ref{fig:finite_tumble_time}b, a finite $k$ optimum appears.
A finite non-zero persistence also appears for diffusive scaling with an added constant,
\textit{i.e.} for Eq. \eqref{eq:tumble_time_scaling} plus a constant.

\section{Chemotactic Strength}
We now ask whether optimality exists for the chemotactic strength parameter $\beta$.
Of course, in models linearised in $\beta$ no such optimality can appear, and we must
seek a different approach.
Averaging over many numerical realisations of the model would allow these effects to be captured,
but a large number of realisations is needed to gain accurate statistics, rendering parameter space exploration hard.
Hence, we begin this section by gaining intuition through a more tractable model,
which gives a good qualitative understanding of the problem.

The crucial insight for this simplified model is that in a constant gradient there is nothing
to distinguish one value of the position variable $x$ from another.
In our full model, the biaser $q(t)$ relaxes to $-\beta$ times the cell's estimate of 
$\cos \theta$ on a time scale $\sim\gamma$.
Such a behaviour can be modelled by the Langevin equation
\begin{align}
& \d q = - \frac{\gamma}{4} \, (q + \beta \cos \theta) \, \d t + \beta \sqrt{2 \sigma} \, \d W~,
\end{align}
where the prefactor of $1/4$ is chosen so that the effective relaxation time matches that of the kernel $\kappa$ used in the full model, and we have introduced a noise term (such a noise term plays no role in the linearised system).
We can specify this system fully through a Fokker-Planck equation for $P(\theta,q,t)$
\begin{equation}
\frac{\partial \hspace{0.05em} P}{\partial t} = \frac{\gamma}{4} \frac{\partial}{\partial q} (q + \beta \cos \theta) \, P + \sigma \beta^2 \, \frac{\partial^2 P}{\partial q^2} + D_r \frac{\partial^2 P}{\partial \theta^2} + \lambda_0 \floor{1 + q} \left[\int \, \frac{e^{k \cos(\theta - \theta')} }{2 \pi I_0(k)} P(\theta') \, \d \theta' - P \right].
\label{eq:fokker_planck_sm}
\end{equation}
The optimal behaviour of the original system is well-captured by this reduced model (Appendix \ref{app:simple}).
Crucially, equation \eqref{eq:fokker_planck_sm} is simple enough to be solved numerically using a hybrid spectral-finite difference method.
Again we find that for all parameters the same efficiency can be obtained for any $k$ by 
a simple rescaling of $\lambda_0$.
We thus set $k=0$ in the remainder of this section without loss of generality.

\begin{figure*}[!t]
\centering
\includegraphics{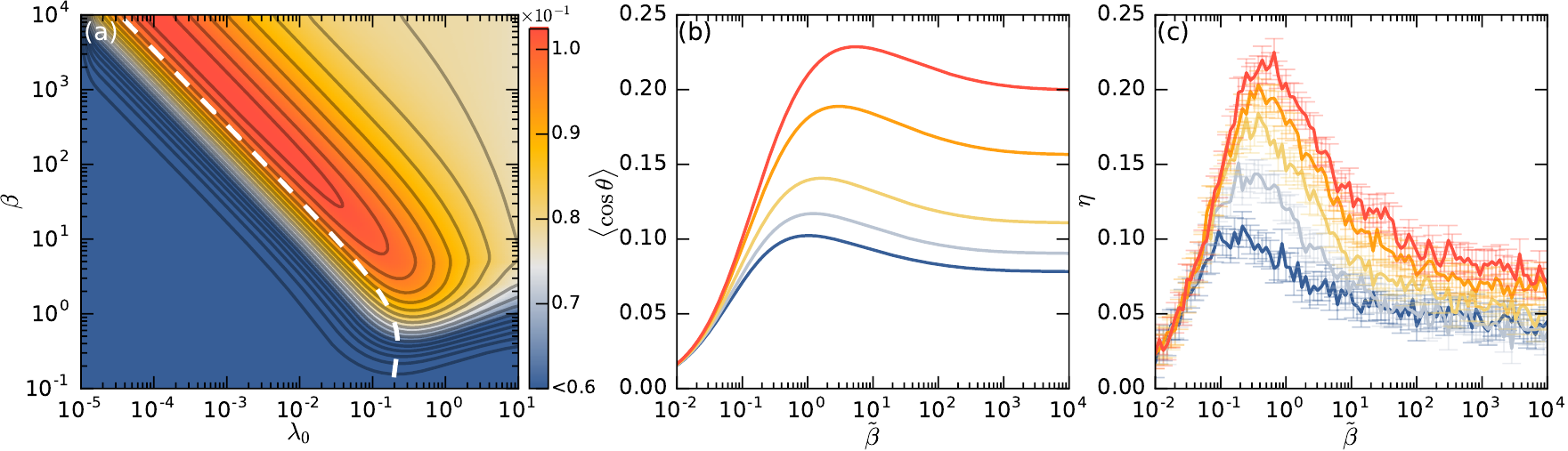}
\caption{Drift efficiency as a function of chemotactic strength. (a) Variations in $\beta$ and $\lambda_0$ reveal a maximum as $\beta \rightarrow \infty$, $\lambda_0 \rightarrow 0$. Dashed curve is analytical approximation to optimum region. (b) Chemotactic drift with modulation of the form $\tilde{\beta} \floor{q}$. Curves vary from $\sigma = 0.15$ (red) to $\sigma = 1.0$ (blue). (c) Full particle simulations with average molecules sensed by cells varying from 0.5 (blue) to 3 (red). 
Common parameters: $D_r = 0.1, \, \beta = 1/2, \, \gamma=1$.}
\label{fig:chemo_strength}
\end{figure*}

Fig. \ref{fig:chemo_strength}a shows the resulting chemotactic drift under variation of the chemotaxis strength $\beta$ and the base tumbling frequency $\lambda_0$.
For a given $\lambda_0$, an optimal chemotactic strength does indeed exist.
Choosing the chemotactic strength too high, evidently, also results in too many tumbles.
Fig. \ref{fig:chemo_strength}a also shows, however, that under variations of both $\beta$ and $\lambda_0$, the optimal is found for $\beta \rightarrow \infty$.
For large $\beta$ the optimum lies on a straight line (power law) relating $\beta$ to $\lambda_0$.

To understand what sets the optimal chemotactic strength,
we seek an analytical approach, but since there is no perturbative small
parameter we examine instead a Fourier-Hermite expansion of the form
\begin{equation}
p(\theta, q) = \sum_{n=0}^N \sum_{m=0}^M a_{nm} \cos(n \theta)  \, H_m(q/\zeta) \, e^{-q^2/\zeta^2},
\end{equation}
where $H_m$ are the Hermite polynomials.
The choice of this expansion arises from the fact that $q$ resembles an Ornstein-Uhlenbeck process,
the solution of which is Gaussian, with a scale $\zeta$, which, for a true
Ornstein-Uhlenbeck process would be $\sim \sqrt{4\sigma/\gamma}$.
Presently, $\cos \theta$ also contributes to variations in $q$, and so $\zeta \sim \sqrt{1 + 4\sigma/\gamma}$.
Here, we truncate at $N=M=1$, which, while yielding numerically inaccurate results nevertheless
reveals the key dynamics.
Higher-order terms can easily be calculated, but the expressions become lengthy.
Exploiting orthogonality, the steady state coefficients $\{ a_{nm} \} = (a_{00}, a_{10}, a_{01}, a_{11})$ are found as the null space of
\begin{align}
\left(
\begin{array}{cccc}
 0 & 0 & 0 & 0  \\
 0 & -\frac{\gamma}{4} & -\frac{\gamma}{8 \zeta} & 0  \\
 0 & 0 & \frac{\lambda_0}{2} \, \text{erfc}( \frac{1}{\zeta \beta }) - \frac{\lambda_0 \zeta \beta}{2 \sqrt{\pi}} \exp(\frac{-1}{\zeta^2 \beta^2}) - \lambda_0 - D_r & -\frac{1}{2} \lambda_0 \zeta \beta (1+ \text{erc}(\frac{1}{\zeta \beta}))   \\
 \frac{-\gamma}{4\beta} & 0 & -\frac{1}{4} \lambda_0 \zeta \beta (1 + \text{erf}(\frac{1}{\zeta \beta})) & \frac{1}{2} \lambda_0  \,\text{erfc}(\frac{1}{\zeta \beta}) - \lambda_0(1 + \frac{\zeta \beta}{\sqrt{\pi}} \exp(\frac{-1}{\zeta^2 \beta^2})
\end{array}
\right),
\end{align}
whence $\langle \cos \theta \rangle = a_{10} / (2 \,a_{00})$.
Optimising this for $\lambda_0$ we obtain the dashed white curve in Fig. \ref{fig:chemo_strength}a.
In the limit $\beta \rightarrow \infty$ this has the form
\begin{equation}
\lambda_0 \sim \frac{D_r (D_r + \gamma/4)}{\sqrt{1 + 4\sigma /\gamma}} \, \frac{1}{\beta}.
\end{equation}
Although the expansion does not quite capture the location of the optimum,
the correct scaling is obtained.
The global optimum is found at $\beta  \rightarrow \infty$ and we learn that $\lambda_0 \beta$ tends to a finite value in that limit.
In detail, the optimisation tries to diminish the base tumbling contribution in the 
expression $\lambda_0 \floor{1 + q}$
and the optimum is found in limit where $\lambda_0 \floor{1 + q} \rightarrow \lambda_0 \floor{q}$.
Explicitly making this substitution in Eq. \eqref{eq:fokker_planck_sm} and defining $\tilde{\beta} = \lambda_0 \beta$,
we obtain a system that has a finite optimal value of chemotactic strength.
This is shown in Fig. \ref{fig:chemo_strength} for various noise strengths $\sigma$.
To verify our conclusions based on this model we turn to the full simulation.
Exactly as in the simplified model, we find optimum behaviour after making the substitution $\lambda_0 \floor{1 + q} \rightarrow \lambda_0 \floor{q}$.
This is shown in Fig. \ref{fig:chemo_strength}c for various levels of chemoattractant concentrations,
confirming our conclusions.

It is perhaps surprising that the optimum is found in this limit, since no modulation of
tumbling frequency then can occur if $q<0$, which is the case when the cell swims just slightly in the correct direction,
and thus in the limit of no noise, the angular distribution will
be governed simply by rotational diffusion on $\theta \in [-\pi/2, \pi/2]$.
In this optimal limit, the cells have minimised the time they spend swimming in any wrong direction,
which, evidently, even though it leads to no active modulation for $q<0$, is also the optimum for the total chemotactic drift.
For stochastic taxis to work, the modulation must necessarily be a monotonically increasing function of $q$.
Strong reaction when swimming in the wrong direction ($q>0$) is thus typically coupled
will smaller reaction when swimming in the correct direction ($q<0$).
Our results show, at least for the presently chosen form of modulation, that with this trade-off
the best choice is to react very strongly when swimming in the wrong direction,
even though this reduces the effectiveness of chemotaxis while swimming in the right direction.

\section{Conclusions}
In this study we have taken an approach to understanding run-and-tumble chemotaxis based on
global parameter optimisation.
For the specific system studied here, cells in constant gradients, we have focused on the
base tumbling frequency $\lambda_0$, tumbling persistence $k$, and chemotactic strength $\beta$ 
as key parameters.  Varying any one parameter alone, there is a unique value that optimises
the chemotactic drift, but when all parameters are free there is a higher-dimensional optimal locus.

In particular, the trade-off in optimality between the base tumbling frequency and 
tumble persistence is ``perfect"
in the sense that any increase, say, in persistence can be countered by an increase in base tumble frequency.
After a persistent tumble, it would seem that the current value of $q$
would stay more relevant than for a full tumble,
indicating that persistence could lead to enhanced chemotaxis.
The intuition behind this argument is based on comparing a \textit{single} full tumble
to a \textit{single} persistent tumble, but a more appropriate comparison 
would be to a \textit{series} of persistent tumbles.
And as evident by our calculations, comparing in this way the argument of preservation of information leads
to similar behaviour for all persistence parameters.
One might also argue for the opposite: 
after a full tumble (or a series of persistent tumbles) there is a high risk that the new direction is wrong.
Therefore, one could argue that keeping $q$ large is a desirable strategy,
since it increases the probability of correcting the tumble quickly.
Our results show that both of these arguments are incorrect.
Although one could imagine a model in which $q$ is explicitly altered after each tumble, say $q \rightarrow a \, q$, the study of this variant would require relaxing the assumption of
fixed $\gamma$ in order to find global optima.  

Introducing a finite tumble time moves the model away from the perfect trade-offs described above.
We have shown that an optimum persistence emerges that depends on the manner in which 
the tumble time depends on reorientation angle.
For ballistic tumbles, zero persistence is optimal,
while continuous tumbling is optimal for diffusive tumbles.
A finite persistence emerges for exponents in-between,
\textit{i.e.} for tumbles that are superdiffusive, but not ballistic.
Such a tumble could simply be a mixture of ballistic and diffusive reorientations that
when taken together have a super-diffusive behaviour.
Diffusive tumbles are easy to generate: a cell can do so by simply not swimming,
and more generally by wiggling its flagella in random directions.
Ballistic tumbles require directed motion of the flagella (even though the actual direction is chosen randomly).
Actual tumbles might be a combination of a fixed tumble time plus a diffusive scaling,
which would favour a finite tumble persistence (non-zero and non-infinite).
One could furthermore imagine minimising tumble times by maintaining a finite swimming speed during tumbles, \textit{e.g.} via polymorphic transformations of the flagella \citep{Goldstein2000}.

In addition to studying the weak chemotaxis limit,
we have also investigated the effects of strong chemotaxis.
While this is, naturally, dependent on the precise functional form chosen for the biasing of tumbles,
we have shown that optima in chemotactic strength can also emerge.
Through an analytical approximation we found that the form $\lambda_0 \floor{1 + q}$ has an optimal value of $\beta$ for constant $\lambda_0$.
Allowing for variations in $\lambda_0$ the optimum shifts to $\beta \rightarrow \infty$,
and instead $\tilde{\beta} = \beta \lambda_0$ as $\beta \rightarrow \infty$ has an optimal value.
This naturally leads to the question of the optimal form of the modulation.
Preliminary results have shown that other simple choices, \textit{e.g.} $\lambda_0 \, e^q$, do not perform better than the form studied here.
In general such problems can be considered partially observable Markov decision processes,
and a potential optimal functional form could be found by methods such as reinforcement learning.
Results from such analysis, however, will probably be strongly dependent on the model setup,
and a form that optimises for drift in constant gradients will not necessarily do well in other gradients.

Comparing to experimental systems, our result that persistence does not have a 
unique optimum when allowing for variations in base
tumble frequency fits well with the variation that exists between species.
The chemotactic strength result that the optimum is found as $\lambda_0 \rightarrow 0$
is a special outcome of maximising the drift velocity in a constant gradient.
In more complex domains, the cells will need to react also to spatial variations (Appendix \ref{app:simple}) and thus need a finite $\lambda_0$ and smaller $\beta$.
Maximising the minimum chemotactic efficiency over many chemoattractant profiles
reveals the experimental values associated with the kernel $\kappa$ and base tumbling frequency \citep{Celani2010}.
A linear approach cannot, however, reveal an optimal chemotactic strength.
An interesting question for future research is thus:
can maximising the minimum chemotactic efficiency over suitably chosen noise models
reveal an optimal finite chemotactic strength?
While difficult to tackle analytically,
numerical methods may be able to answer such questions.

\section*{Acknowledgments}
It is a pleasure to dedicate this work to the memory of John Blake, whose 
impact on the mathematics of microorganism locomotion was so profound.  
This work was supported in part by the EPSRC and St. John's College, Cambridge (JBK),
and by an Established Career Fellowship from the EPSRC (REG).

\appendix
\section{Linearised drift}
\label{app:lindrift}
To find the drift $\langle \cos \theta \rangle$ linearised in $\beta$, we multiply 
\begin{align}
\frac{\partial P}{\partial t} + v \cos \theta \, \frac{\partial P}{\partial x}  = & \, D_r \frac{\partial^2 P}{\partial \theta^2} + \lambda_0 \floor{1 + q(t)} \left[\int \, \frac{e^{k \cos(\theta - \theta')} }{2 \pi I_0(k)} P(\theta') \, \d \theta' - P \right] \\ \nonumber
&- \sum_{j} \partial_{m_j} [ \delta_{j,0} \, c(x) \, + j m_{j-1} - \gamma m_j ] \, P,
\end{align}
by $\cos \theta$, whereafter integration yields
\begin{equation}
\partial_t \langle \cos \theta \rangle = -D_r \langle \cos \theta \rangle - \lambda_0 \left( 1 - \frac{I_1(k)}{I_0(k)} \right) \left( \langle \cos \theta \rangle + \langle q \cos \theta \rangle \right),
\end{equation}
using
\begin{equation}
\int e^{k \cos(\theta - \theta')} \cos \theta\, \d \theta = 2 \pi I_1(k) \cos \theta'.
\end{equation}
Since $q(t) = \frac{\beta \gamma^2}{\alpha \, v} (\gamma^2 m_2/2 - \gamma m_1)$ we continue, neglecting quadratic terms
\begin{align}
\partial_t \langle m_0 \cos \theta \rangle &= -(D_r + \lambda_k + \gamma) \langle m_0 \cos \theta \rangle + \alpha \langle x \cos \theta \rangle ,    \\
\partial_t \langle m_1 \cos \theta \rangle &= -(D_r + \lambda_k + \gamma) \langle m_1 \cos \theta \rangle +  \langle m_0 \cos \theta \rangle,  \\
\partial_t \langle m_2 \cos \theta \rangle &= -(D_r + \lambda_k + \gamma) \langle m_2 \cos \theta \rangle +  2 \langle m_1 \cos \theta \rangle ,   \\
\partial_t \langle x \cos \theta \rangle &= \frac{v}{2} - (D_r + \lambda_k)  \langle x \cos \theta \rangle.
\end{align}
Solving these equations for the steady state, one finds the result of the main text.

\section{Linearised with mean tumble angle}
\label{app:meantumble}
The reorientation distribution is now
\begin{equation}
g_k(\theta, \theta') = \frac{1}{4 \pi I_0(k)} \left( e^{k \cos(\theta - \theta'-\mu)} + e^{k \cos(\theta - \theta' + \mu)} \right)
\end{equation}
such that
\begin{equation}
\partial_t \langle \cos \theta \rangle = -D_r \langle \cos \theta \rangle - \lambda_0 \left( 1 - \frac{I_1(k)}{I_0(k)}  \cos \mu \right) \left( \langle \cos \theta \rangle + \langle q \cos \theta \rangle \right),
\end{equation}
where we used that $\cos \mu$ is an even function and $\sin \mu$ odd.
So having a finite $\mu$ corresponds to changing the persistence.
At precisely $\mu = \pm \pi/2$, persistence no longer changes the behaviour.

\section{Linearised with mean tumble angle --- biased direction}
\label{app:biastumble}
Here we take
\begin{equation}
g_k(\theta, \theta') = \frac{1}{2 \pi I_0(k)} e^{k \cos(\theta - \theta'-\mu)}.
\end{equation}
We now have
\begin{align} \nonumber
\partial_t \langle \cos \theta \rangle = &-D_r \langle \cos \theta \rangle - \lambda_0 \left( 1 - \frac{I_1(k)}{I_0(k)} \cos \mu \right)  \left( \langle \cos \theta \rangle + \langle q \cos \theta \rangle \right) \\
&-\lambda_0 \frac{I_1(k)}{I_0(k)} \sin \mu \left( \langle \sin \theta \rangle + \langle q \sin \theta \rangle \right).
\end{align}
And then
\begin{align} \nonumber
\partial_t \langle m_0 \cos \theta \rangle &= -D_r \langle m_0 \cos \theta \rangle - \lambda_0 \left( 1 - \frac{I_1(k)}{I_0(k)}\cos \mu \right)  \langle m_0 \cos \theta \rangle \\
&\quad  - \lambda_0\frac{I_1(k)}{I_0(k)} \sin \mu \langle m_0 \sin \theta \rangle + \alpha \langle x \cos \theta \rangle  - \gamma \langle m_0 \cos \theta \rangle,    \\ \nonumber
\partial_t \langle m_1 \cos \theta \rangle &= -D_r \langle m_1 \cos \theta \rangle - \lambda_0 \left( 1 - \frac{I_1(k)}{I_0(k)} \cos \mu \right)  \langle m_1 \cos \theta \rangle \\
& \quad - \lambda_0\frac{I_1(k)}{I_0(k)} \sin \mu \langle m_1 \sin \theta \rangle +  \langle m_0 \cos \theta \rangle  - \gamma \langle m_0 \cos \theta \rangle,  \\ \nonumber
\partial_t \langle m_2 \cos \theta \rangle &= -D_r \langle m_2 \cos \theta \rangle - \lambda_0 \left( 1 - \frac{I_1(k)}{I_0(k)} \cos \mu \right)  \langle m_2 \cos \theta \rangle \\
& \quad - \lambda_0\frac{I_1(k)}{I_0(k)} \sin \mu \langle m_2 \sin \theta \rangle +  2 \langle m_1 \cos \theta \rangle  - \gamma \langle m_2 \cos \theta \rangle,   \\ \nonumber
\partial_t \langle x \cos \theta \rangle &= \frac{v}{2} -D_r \langle x \cos \theta \rangle - \lambda_0 \left( 1 - \frac{I_1(k)}{I_0(k)} \cos \mu \right)  \langle x \cos \theta \rangle \\
& \quad - \lambda_0\frac{I_1(k)}{I_0(k)} \sin \mu \langle x \sin \theta \rangle
\end{align}
and similarly for the $\sin \theta$ terms, except no $v/2$ term appears in the equation for $\partial_t \langle x \sin \theta \rangle$.

This can be solved for the steady solution of $\langle \cos \theta \rangle$, but the expression is quite lengthy.
Analysing it, we find that the optimum is found for $k \rightarrow \infty$.
Taking this limit we find
\begin{align}
\langle \cos \theta \rangle \rightarrow & \beta \gamma ^3 \lambda_0 \bigg[ \lambda_0 ^2 (\cos (2 \mu ) (3 (\gamma +D_r+\lambda_0 ) (\gamma +2 (D_r+\lambda_0 ))+\lambda_0 (3 \gamma +4 (D_r+\lambda_0 ))) \\ \nonumber
&+\lambda_0  (\cos (3 \mu ) (-3 \gamma -4 D_r-5 \lambda_0 )+\lambda_0  \cos (4 \mu )))-\cos \mu (\gamma +D_r+\lambda_0 ) \big(\lambda_0 ^2 (10 \gamma +17 D_r) \\ \nonumber
&+\lambda_0  (\gamma +D_r) (2 \gamma +7 D_r)+D_r (\gamma +D_r)^2+11 \lambda_0 ^3\big)+(\gamma +D_r+\lambda_0 )^2 ((D_r+\lambda_0 ) (\gamma +D_r+\lambda_0 ) \\ \nonumber
&+\lambda_0  (\gamma +4 (D_r+\lambda_0 ))) \bigg] \bigg/ \bigg[ 2 \big(-2 \lambda_0  (D_r+\lambda_0 ) \cos \mu +(D_r+\lambda_0 )^2+\lambda_0 ^2\big) \\ \nonumber
& \big(-2 \lambda_0  \cos \mu (\gamma +D_r+\lambda_0 )+(\gamma +D_r+\lambda_0 )^2+\lambda_0 ^2\big)^3 \bigg].
\end{align}
The optimum is found at $\mu \rightarrow 0$, $\lambda_0 \rightarrow \infty$, keeping $\lambda_0 \mu$ constant.
The motion is thus continuously rotating cells,
where the rotation speed is modulated by the chemoattractants.

\section{Persistence in 3D}
\label{app:3d}
The linearised calculation is very similar in 3D.
Defining $p$ as a unit vector in the swimming direction,
we can write the Fokker-Planck equation with von-Mises tumbles as
\begin{align}
\frac{\partial P(t,x,p,m)}{\partial t} + v \, p_x \, \frac{\partial P}{\partial x}  = & \, D_r \nabla_p^2 P + \lambda_0 \floor{1 + q(t)} \left[\int \, \frac{k \, e^{k \, p \cdot p'} }{4 \pi \sinh k} P(p') \, \d \Omega' - P \right] \\ \nonumber
&- \sum_{j} \partial_{m_j} [ \delta_{j,0} \, c(x) \, + j m_{j-1} - \gamma m_j ] \, P,
\end{align}
where $\nabla_p^2$ is the angular Laplacian.

This leads to 
\begin{equation}
\partial_t \langle p_x \rangle = -2 D_r \langle p_x \rangle - \lambda_0 \left[1 + \frac{1}{k} - \frac{1}{\tanh k} \right] \left( \langle p_x \rangle + \langle q \, p_x \rangle \right),
\end{equation}
showing an only slightly altered persistence modification to $\lambda_k$ compared to 2D,
and thus leading to similar conclusions as in 2D.

\section{Simplified effective model}
\label{app:simple}
\begin{figure}[t]
\centering
\includegraphics{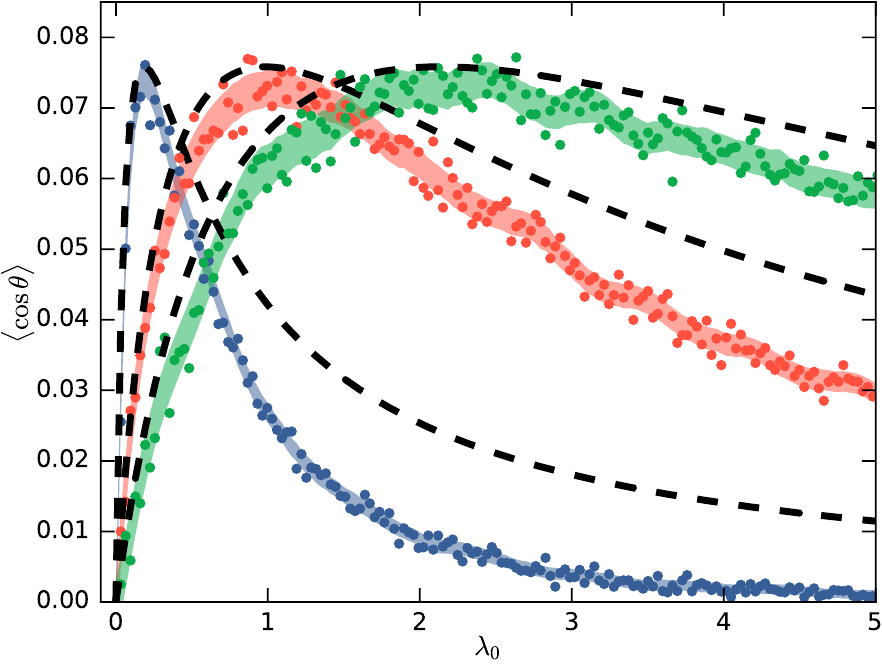}
\caption{Similar to Fig. \ref{fig:vary_k}, but with theoretical curves (dashed) 
obtained from the analysis of Eq. \eqref{eq:fokker_planck_sm}.}
\label{fig:fokker_planck_sm}
\end{figure}

Fig. \ref{fig:fokker_planck_sm} shows the performance of the simplified model of Eq. \eqref{eq:fokker_planck_sm} compared to the simulations shown in Fig. \ref{fig:vary_k}.

\section{Run-and-tumble in discrete 1D}
\label{app:1d}
In this section we study run-and-tumble in one spatial dimension.
Such simplifications have proven to yield much insight in the case of no noise \citep{Tranquillo1989}.
For the present purpose we will consider the case of a very noisy signal.
To simplify further we put the cells and chemoattractants on an equilateral grid.
We assume that each measurement carries the same error $\sigma$.
In reality, the measurement of a concentration $c$ has an error $ \propto \sqrt{c}$,
but such effects are expected to be second order.
Thus at each grid position $i$, the cell measures a concentration $c_i \sim \mathcal{N}(c(\x_i), \sigma^2)$, where $c(\x_i)$ is the time-average of the signal at that position.

In particular, a cell will swim $n$ lattice points and calculate $Q = \sum_j k_j c_{i(j)}$,
where $k_j$ is some kernel,
reminiscent of the continuous kernel used in the main text.
In the simplest model, if $Q>0$ the cell will keep
going in the same direction, but turn if $Q<0$.
We begin by determining the optimal $\{ k_j \}$.

\subsection{Optimal kernel}
We determine $\{k_j \}$ in such a way that $Q$ makes the best estimate of a constant gradient.
Thus we consider $c_i = \mathcal{N}(c_0 + \alpha i), \, \sigma^2)$.
Then $Q = \sum k_j c_j$ is an unbiased estimator of $\alpha$ if
\begin{equation}
\langle Q \rangle = \sum k_j \langle c_j \rangle = \sum k_j (c_0 + \alpha j) = c_0 \sum k_j + \alpha \sum k_j j = \alpha,
\end{equation}
so we must require $\sum k_j = 0$ and $\sum k_j j = 1$.
The variance becomes
\begin{equation}
\text{Var} Q = \langle Q^2 \rangle - \langle Q \rangle^2 = \sigma^2 \sum k_j^2.
\end{equation}
Writing $Q = \sum k_i c_i$, minimising the variance subject to the unbiased estimation yields
\begin{equation}
k_i = \frac{6(2i - n_1 - n_2)}{(n_2 - n_1+2)(n_2 - n_1 + 1)(n_2-n1)}
\end{equation}
for a measurement on $[n_1, n_2]$.
Thus
\begin{equation}
k_j = \frac{6(2j - n)}{(n+2)(n + 1)n},
\end{equation}
where $n = n_2 - n_1$.
Assuming a Gaussian distribution we thus have
\begin{equation}
Q \sim \mathcal{N} \! \left(\alpha, \sigma^2 \sum k_j^2 \right) = \mathcal{N} \! \left(\alpha, \frac{12 \sigma^2}{(n+2)(n+1)n}\right) \equiv  \mathcal{N}(\alpha, \sigma_n^2).
\end{equation}

\subsection{Chemotaxis in a constant gradient}
The probability that $Q<0$ after a swim of $n$ lattice points follows a geometric distribution
with parameter
\begin{equation}
q = \frac{1}{\sqrt{2 \pi \sigma^2}} \int_{-\infty}^0 e^{-(x - \alpha)^2 / 2 \sigma_n^2} \, \d x =  \frac{1}{2} \, \text{erfc} \left( \frac{\alpha}{\sqrt{2 \sigma_n^2}} \right).
\label{eq:q}
\end{equation}
After a run left and right, the cell will have travelled on average
\begin{equation}
 \frac{n}{q} - \frac{n}{1-q},
\end{equation}
while in that time it could have travelled on average the distance
\begin{equation}
\frac{n}{q} + \frac{n}{1-q}.
\end{equation}
The efficiency is thus $\eta = 1-2q$, which is maximised for $q \rightarrow 0$, corresponding to $n \rightarrow \infty$.
This is in the absence of spatial variations and diffusion effects.

\subsection{Effective rotational diffusion}
\begin{figure}[t]
	\centering
    \includegraphics{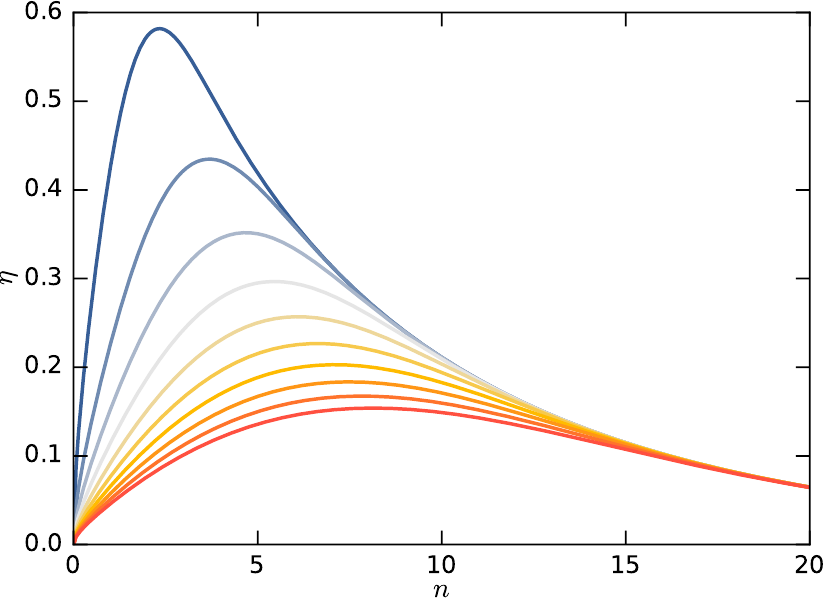}
    \caption{Efficiency $\eta$ as a function of $n$ as in Eq. \eqref{eq:1ddr}. Noise varies from $\sigma/\alpha = 1$ (blue) to $\sigma/\alpha = 10$ (red). $D_r = 0.1$.} \label{fig:dr1d}
\end{figure}

We now add the feature that after each jump
the particle will flip either because $Q<0$ or another process $R<0$, which has parameter $D_r$ for a single jump.
The probability $R<0$ after $n$ jumps will thus be $r = 1-(1-D_r)^n$.
Thus the probability of a turn after the $n$ jumps is $\tilde{q} = q + r - qr$ when going right, and when going left $\tilde{p} = (1-q) + r - (1-qr)$.
Calculating the efficiency we thus find
\begin{equation}
\eta = \frac{(1-2q) \, (1-D_r)^n}{2 - (1-D_r)^n},
\label{eq:1ddr}
\end{equation}
where $q$ is defined as in Eq. \eqref{eq:q}.
This defines an optimal $n$ as shown in Fig. \ref{fig:dr1d}.
The emergence of an optimal $n$ corresponds to the emergence of an optimum $\gamma$ in the main text.

\subsection{Chemotaxis in a spatially varying gradient}
Spatial variation from linear concentration profiles can also affect the optimal choice of $n$.
Consider cells swimming in a gradient being held to a fixed value $c_0 > 0$ at the origin.
The diffusion equation allows for linear steady state solutions.
We thus consider 1D swimmers in
\begin{equation}
c(x) = c_0 - \alpha |x|.
\end{equation}
We again discretise space and allow the cells to choose an $n$,
the number of lattice points to swim before making a decision on whether to change direction.
This $n$ determines $\sigma_n$ and thus $q$.
This also makes the cells only visit sites that are multiples of $n$ and we thus reindex by that.
We further note that a state moving right at position $i$ is by symmetry the same as
moving left at site $-i$.
We exploit this symmetry and consider only $i \geq 0$.
Our states are then called
\begin{equation}
(i, \, s) \in \mathbb{N}_0 \times\{+,-\} = \left\{ \begin{matrix}
(0,+), & (0,-),  & (1,+),  &  (1,-),  &  (2,+),  &  (2,-), & \cdots  
\end{matrix} \right\}.
\end{equation}

\begin{figure}[tb]
	\centering
    \includegraphics{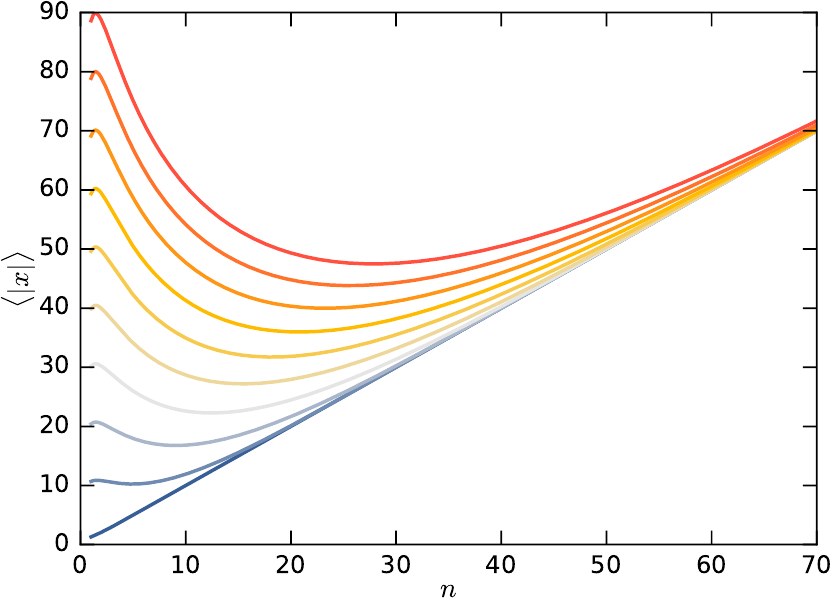}
    \caption{Average $|x|$ for as a function of $n$, the number of jumps before deciding to change direction, as described by Eq. \eqref{eq:meanx}. Chemotaxis is optimal when $\langle |x| \rangle$ is minimised, since the chemoattractant concentration decreases from $x=0$. Parameters as in Fig. \ref{fig:dr1d}.} \label{fig:meanx}
\end{figure}

The jumps form an infinite dimensional Markov chain with transition matrix
\begin{equation}
\mathbb{T} = \begin{pmatrix}
\0 & \0 & \0 & p & \0 & \0 & \0 & \0 & \0 & \cdots \\
\0 & \0 & \0 & q & \0 & \0 & \0 & \0 & \0 & \cdots \\
p & q & \0 & \0 & \0 & p & \0 & \0 & \0 & \cdots \\
q & p & \0 & \0 & \0 & q & \0 & \0 & \0 & \cdots \\
\0 & \0 & p & \0 & \0 & \0 & \0 & p & \0 & \cdots \\
\0 & \0 & q & \0 & \0 & \0 & \0 & q & \0 & \cdots \\
\0 & \0 & \0 & \0 & p & \0 & \0 & \0 & \0 & \cdots \\
\0 & \0 & \0 & \0 & q & \0 & \0 & \0 & \0 & \cdots \\
\0 & \0 & \0 & \0 & \0 & \0 & p & \0 & \0 & \cdots \\
\0 & \0 & \0 & \0 & \0 & \0 & q & \0 & \0 & \cdots \\
\vdots & \vdots & \vdots & \vdots & \vdots & \vdots & \vdots & \vdots & \vdots & \ddots
\end{pmatrix},
\end{equation}
where $p = 1-q$.
The steady state distribution is found by solving
\begin{equation}
{\bf p } = \mathbb{T} \, {\bf p }.
\end{equation}
To solve this infinite set of equations, we truncate in an appropriate manner
at $m$ equations and then let $m \rightarrow \infty$.
We assume $m$ even and to conserve probability, for finite $m$ set $\mathbb{T}_{m-1,m-1} = p$ and $\mathbb{T}_{m,m-1} = q$.
This leads to
\begin{equation}
p_{i, s} = \frac{1}{A} \cdot
\begin{cases} 
  \frac{q^k}{p^{k-2}(1-2q+2q^2)} & (0,+) \\
  \frac{q^{k+1}}{p^{k-1}(1-2q+2q^2)} & (0,-) \\
  \frac{q^{k-i}}{p^{k-i}} & (i \geq 1, +) \\
  \frac{q^k}{p^{k-2}(1-2q+2q^2)} & (1,-) \\
  \frac{q^{k-i+1}}{p^{k-i+1}} & (i \geq 2, -)
\end{cases}
\end{equation}
where $k=m/2-2$ and $A$ is a normalisation constant determined by
\begin{align}
\sum_{i,s}  \, p_{i,s} = 1.
\end{align}
After a long calculation we find (for $q>1/2$) in the limit $m \rightarrow \infty$,
\begin{equation}
\langle |x| \rangle = \frac{\Delta x \, n}{2} \left(2 - \frac{1}{q(n)} + \frac{1}{2q(n)-1} \right).
\label{eq:meanx}
\end{equation}
Fig. \ref{fig:meanx} shows a minimum appearing for large noise.
Thus we see that an optimal measurement distance must also be balanced with potential spatial variations,
not just with rotational diffusion.

\bibliographystyle{imamat}

\begin{thebibliography}{}

\bibitem[Friedrich and Juicher, 2007]{Friedrich2007}
Friedrich, B. M. \& Julicher, F.
\newblock {Chemotaxis of sperm cells}.
\newblock {\em Proceedings of the National Academy of Sciences of the United
  States of America}, 2007.
  
\bibitem[Jikeli \textit{et~al.}, 2015]{Jikeli2015}
Jikeli, J. F., Alvarez, L., Friedrich, B. M. Wilson, L. G., Pascal, R., Colin, R., Pichlo, M., Rennhack, A.,Brenker, C. \& Kaupp, U. B.
\newblock {Sperm navigation along helical paths in 3D chemoattractant
  landscapes}.
\newblock {\em Nature communications}, 6:7985, 2015.

\bibitem[Yoshimura and Kamiya, 2001]{Yoshimura2001}
Yoshimura, K. \& Kamiya, R.
\newblock {The sensitivity of chlamydomonas photoreceptor is optimized for the
  frequency of cell body rotation.}
\newblock {\em Plant {\&} cell physiology}, 42(6):665--672, 2001.

\bibitem[Drescher \textit{et~al.}, 2010]{Drescher2010}
Drescher, K., Goldstein, R. E. \& Tuval, I.
\newblock {Fidelity of adaptive phototaxis.}
\newblock {\em Proceedings of the National Academy of Sciences of the United
  States of America}, 107(25):11171--6, 2010.
  
\bibitem[Polin \textit{et~al.}, 2009]{Polin2009}
Polin, M., Tuval, I., Drescher, K., Gollub, J. P. \& Goldstein, R. E.
\newblock {Chlamydomonas Swims With Two `Gears' in a Eukaryotic Version of Run-and-Tumble Locomotion.}
\newblock {\em Science}, 325 487-490, 2009.

\bibitem[Bonner and Savage, 1947]{Bonner1947}
Bonner, J. T. \& Savage, L.
\newblock {Evidence for the formation of cell aggregates by chemotaxis in the development of the slime mold Dictyostelium discoideum}.
\newblock {\em Journal of Experimental Zoology}, 106(1), 1947.

\bibitem[Berg and Brown, 1972]{Berg1972}
Berg, H. C. \& Brown, D. A.
\newblock {Chemotaxis in Escherichia coli analysed by three-dimensional
  tracking}.
\newblock {\em Nature}, 1972.

\bibitem[Kirkegaard \textit{et~al.}, 2016a]{Kirkegaard2016a}
Kirkegaard, J. B., Bouillant, A., Marron, A. O., Leptos, K. C. \&
  Goldstein, R. E.
\newblock {Aerotaxis in the Closest Relatives of Animals}.
\newblock {\em eLife}, e18109, 2016.

\bibitem[Kirkegaard \textit{et~al.}, 2016b]{Kirkegaard2016b}
Kirkegaard, J. B., Marron, A. O. \&
  Goldstein, R. E.
\newblock {Motility of Colonial Choanoflagellates and the Statistics of Aggregate Random Walkers}.
\newblock {\em Physical Review Letters}, 116 038102, 2016.

\bibitem[Tindall \textit{et~al.}, 2008a]{Tindall2008a}
Tindall, M. J, Porter, S. L., Maini, P. K, Gaglia, G. \& Armitage, J. P.
\newblock {Overview of mathematical approaches used to model bacterial
  chemotaxis I: The single cell}.
\newblock {\em Bulletin of Mathematical Biology}, 70(6):1525--1569, 2008.

\bibitem[Tindall \textit{et~al.}, 2008b]{Tindall2008b}
Tindall, M. J, Maini, P. K, Porter, S. L. \& Armitage, J. P.
\newblock {Overview of mathematical approaches used to model bacterial
  chemotaxis II: Bacterial populations}.
\newblock {\em Bulletin of Mathematical Biology}, 70(6):1570--1607, 2008.

\bibitem[Segall \textit{et~al.}, 1986]{Segall1986}
Segall, J. E., Block, S. M. \& Berg, H. C.
\newblock {Temporal comparisons in bacterial chemotaxis.}
\newblock {\em Proceedings of the National Academy of Sciences of the United
  States of America}, 83(23):8987--8991, 1986.

\bibitem[Celani and Vergassola, 2010]{Celani2010}
Celani, A. \& Vergassola, M.
\newblock {Bacterial strategies for chemotaxis response}.
\newblock {\em Proceedings of the National Academy of Sciences},
  107(4):1391--1396, 2010.

\bibitem[Mora and Wingreen, 2010]{Mora2010}
Mora, T. \& Wingreen, N. S.
\newblock {Limits of sensing temporal concentration changes by single cells}.
\newblock {\em Physical Review Letters}, 104(24):1--4, 2010.

\bibitem[Hein \textit{et~al.}, 2016]{Hein2016}
Hein, A. M., Brumley, D. R., Carrara, F., Stocker, R. \& Levin, S. A.
\newblock {Physical limits on bacterial navigation in dynamic environments.}
\newblock {\em Journal of the Royal Society},
  13(114):20150844, 2016.

\bibitem[Locsei, 2007]{Locsei2007}
Locsei, J. T.
\newblock {Persistence of direction increases the drift velocity of run and
  tumble chemotaxis}.
\newblock {\em Journal of Mathematical Biology}, 55(1):41--60, 2007.

\bibitem[Locsei and Pedley, 2009]{Locsei2009}
Locsei, J. T. \& Pedley, T. J.
\newblock {Run and tumble chemotaxis in a shear flow: The effect of temporal
  comparisons, persistence, rotational diffusion \& cell shape}.
\newblock {\em Bulletin of Mathematical Biology}, 71(5):1089--1116, 2009.

\bibitem[Reneaux and Gopalakrishnan, 2010]{Reneaux2010}
Reneaux, M. \& Gopalakrishnan, M.
\newblock {Theoretical results for chemotactic response and drift of E. coli in
  a weak attractant gradient}.
\newblock {\em Journal of Theoretical Biology}, 266(1):99--106, 2010.

\bibitem[Mortimer \textit{et~al.}, 2011]{Mortimer2011}
Mortimer, D., Dayan, P., Burrage, K. \& Goodhill, G. J.
\newblock {Bayes-optimal chemotaxis.}
\newblock {\em Neural computation}, 23(2):336--373, 2011.

\bibitem[Nicolau \textit{et~al.}, 2009]{Nicolau2009}
Nicolau, D. V., Armitage, J. P. \& Maini, P. K.
\newblock {Directional persistence and the optimality of run-and-tumble
  chemotaxis}.
\newblock {\em Computational Biology and Chemistry}, 33(4):269--274, 2009.

\bibitem[Peaudecerf and Goldstein, 2015]{Peaudecerf2015}
Peaudecerf, F. J. \& Goldstein, R. E.
\newblock {Feeding ducks, bacterial chemotaxis \& the Gini index}.
\newblock {\em Phys. Rev. E}, 92(022701), 2015.

\bibitem[Berg and Purcell, 1977]{BergPurcell}
Berg, H. C. \& Purcell, E. M.
\newblock {Physics of Chemoreception}
\newblock {\em Biophysical Journal} 20, 1977

\bibitem[Goldstein \textit{et~al.}, 2000]{Goldstein2000}
Goldstein, R. E., Goriely, A., Huber, G. \& Wolgemuth, C. W.
\newblock {Bistable Helices}.
\newblock {\em Physical Review Letters}, 84(7):1631--1634, 2000.

\bibitem[Rivero \textit{et~al.}, 1989]{Tranquillo1989}
Rivero, M. A., Tranquillo, R. T., Buettner, H. M. \& Lauffenburger, D. A.
\newblock {Transport models for chemotactic cell populations based on
  individual cell behavior}.
\newblock {\em Chemical Engineering Science}, 44(12):2881--2897, 1989.


\end{thebibliography}

\end{document}